\documentclass[traditabstract]{aa} 
\usepackage{graphicx}
\usepackage{comment}

\usepackage{natbib} 
\bibpunct{(}{)}{;}{a}{}{,} 
\usepackage{multirow,bigdelim}
\usepackage{rotating}
\usepackage[title]{appendix}

\usepackage{subfigure}

\def\lsim{ \lower .75ex\hbox{$\sim$} \llap{\raise .27ex \hbox{$<$}} }
\def\gsim{ \lower .75ex \hbox{$\sim$} \llap{\raise .27ex \hbox{$>$}} }

\newcommand{\bi}{\begin{itemize}}
\newcommand{\ei}{\end{itemize}}

\def\deg{^\circ}

\newcommand{\virg}[1]{``#1"}
\def \ergsc{\hbox{erg s$^{-1}$ cm$^{-2}$}}

\def \chirid{$\chi ^{2}_{\nu}$}
\newcommand{\I }{{\it I}}
\newcommand{\Q }{{\it Q}}
\newcommand{\U}{{\it U}}
\newcommand{\mdp}{{\it MDP}}

\begin{document}

\title{Testing particle acceleration models for BL Lac jets with the {\it Imaging X-ray Polarimetry Explorer}}
\author{L. Di Gesu \inst{1}
\and F. Tavecchio \inst{2}
\and I. Donnarumma \inst{1}
\and A. Marscher \inst{3}
\and M. Pesce-Rollins \inst{4}
\and M. Landoni \inst{2}}
\institute
{Italian Space Agency (ASI), Via del Politecnico snc, 00133, Roma, Italy
\and
INAF – Osservatorio Astronomico di Brera, via E. Bianchi 46, I–23807 Merate, Italy
\and
Institute for Astrophysical Research, Boston University, Boston, MA 02215, USA
\and
Istituto Nazionale di Fisica Nucleare, Sezione di Pisa, I-56127 Pisa, Italy}
\date{}
\abstract{
%
%
Mechanisms such as shock acceleration, magnetic reconnection in a kink unstable jet, and extreme turbulence in the jet flow are all expected to produce a distinctive time variability pattern of the X-ray polarization properties of high synchrotron peak blazars (HSP). 
%
%
To determine whether  the recently launched  Imaging X-ray Polarimetry Explorer (IXPE) can follow the polarization
variations induced by different particle acceleration mechanisms in blazar jets,
we simulated observations of an HSP blazar variable in terms of the polarization degree and angle according to  theoretical predictions.
%
%
We used the Monte Carlo tool {\tt ixpeobssim} to create realistic
IXPE data products for each model and for three values of flux (i.e., 1, 5, and 10 $\times 10^{-10}$ \ergsc). We generated simulated light curves of the polarization degree and angle by time-slicing the simulated data into arbitrary short time bins. We used an $\chi^2$ test to assess the performance of the observations in detecting the time variability  of the polarization properties.\\
%
%
In all  cases, even when the light curves are diluted in an individual time bin, some degree of polarization is still measurable with IXPE. A series of $\sim$10 ks long observations permitted IXPE to follow the time variability of the polarization degree in the case of the shock acceleration model. In the case of the magnetic reconnection model, the nominal injected model provides the best fit of the simulated IXPE data for time bins of $\sim$5-10 ks, depending on the tested flux level. For the TEMZ model, shorter time slices of $\sim$0.5 ks are needed for obtaining a formally good fit of the simulated IXPE data with the injected model. On the other hand, we find that a fit with a constant model provides a $\chi^2$ lower than the fit with the nominal injected model when using time slices of $\sim$20 ks, $\sim$60/70 ks, and $\sim$5 ks for the case of the shock acceleration, magnetic reconnection, and TEMZ model, respectively. \\
%
%
In conclusion, provided that the statistics
of the observation allows for the slicing of the data
in adequately short time bins, IXPE observations of an HSP blazar at a typical flux level can detect the time variability predicted by popular models for particle acceleration in jets. IXPE observations of HSP blazars are a useful tool for addressing the issue of particle acceleration in blazar jets.}
\titlerunning{Testing particle acceleration models for BL Lac jets with IXPE} 
\authorrunning{Di Gesu et al.}
\maketitle
%
%
\section{Introduction}

Efficient particle acceleration, identified through intense non-thermal emission, is a prominent property of relativistic jets \citep[e.g.,][]{blandford19}. The powerful gamma-ray emission of blazars (characterized by a jet closely oriented toward the Earth) indicates that particles (leptons and, possibly, nuclei) can be pushed to energies as high as several tens of TeV. However, current observations are incapable of uniquely identifying the process(es)  particles go through to reach such high energies. 

Diffusive shock acceleration (DSA) is classically considered the main mechanism for particle acceleration in several astrophysical environments \citep{blandford87}. For relativistic jets, shocks have long been considered the main structures where particles can be accelerated and emission is produced \citep[e.g.,][]{marscher78, blandford79}. Shocks naturally arise in the supersonic flows that characterize relativistic jets. Internal shocks are the result of unsteady flows (e.g., with a variable bulk Lorentz factor, \citealt{spada2001}), while reconfinement shocks form when an expanding outflow recollimates under the effect of the pressure exerted by an external medium \citep[e.g.,][]{daly1988}. In a real jet, both kinds of shocks are likely to occur, possibly at different distances from the central engine. Moreover, turbulence is expected to develop in the downstream region of the shock, affecting both the acceleration and emission of high-energy particles.

Recent studies have pointed out that for jets with high magnetization, DSA is relatively inefficient and works only for mildly relativistic shocks with configurations in which the magnetic field lines in the upstream flow are nearly parallel to the shock normal (i.e., "parallel"\ shocks) \citep{sironi15}. This supports the view that for large jet magnetization, particles are accelerated by the direct release of magnetic energy through reconnection of field lines. This scenario is consistent with the results of MHD simulations, suggesting that jets start out as magnetically dominated outflows in which the magnetic energy is progressively converted to kinetic energy while the jet accelerates \cite[e.g.,][]{komissarov07,tchek09}. Magnetically dominated jets are intrinsically unstable, which is due, in particular, to the well studied current-driven kink instability \cite[e.g.,][]{bodo21} whose non-linear stages create the conditions for efficient dissipation of magnetic energy through magnetic reconnection.  Particle-in-cell simulations have shown that current sheets formed in highly magnetized plasmas are sites of fast, relativistic reconnection that sustains efficient acceleration of particles with non-thermal (power law) distributions \citep[e.g.,][]{zenitani01, sironi14, guo15, werner16}.

Simulations and theoretical studies may set out to explore the landscape of potential physical processes, but only observational evidence can determine which mechanism(s) is responsible for particle acceleration in jets. In this respect, polarimetry is a powerful tool to probe  magnetic field geometries and particle acceleration \citep[e.g.,][]{angel80, blandford19}. 
The launch of the Imaging X-ray Polarimetry Explorer (IXPE, \citealt{ixpe21}) satellite promises to revolutionize the  study of the polarimetric properties of relativistic jets in blazars and other cosmic sources. Indeed, thanks
to the three Gas Pixel Detector Units (DU) in its focal plane, IXPE will allow, for the first time, for a measurement of the X-ray polarization (i.e., in the 2.0-8.0 keV band) of extragalactic sources such as blazars. In particular, IXPE observations of blazars of the HSP type (in which the synchrotron emission by relativistic  electrons extends up to the X-ray band) will be instrumental in the investigations of the mechanism that accelerates electrons up to multi-TeV energies \citep[e.g.,][]{tavecchio21}. In this paper, we present a first set of simulations aimed at studying the possibility to effectively test and possibly constrain current models for shocks (with different conditions), magnetic reconnection acceleration, and turbulence in the jet. In particular, we will  take advantage of synthetic light curves, including polarization (i.e., polarization degree, $P,$ and electric vector position angle, hereafter denoted as the  \virg{polarization angle} $\theta$), presented by \citet{tavecchio20}, \citet{bodo21}, and \citet{marscher21} and of Montecarlo simulations of realistic IXPE data products.\\
The paper is structured as follows: 
in Sect. 2 we briefly present the models used for simulations,
while in in Sect. 3 we describe the simulation procedure. Finally,
in Sect. 3 we extract the results of the simulations and in Sect. 4
we discuss our findings.

\section{The models}

\subsection{Shock acceleration with self-produced fields}

It has been widely established that DSA requires the presence of a sufficient level of magnetic turbulence in the flow to ensure the effective scattering of accelerating particles. It is broadly accepted that the magnetic field is produced by the particles themselves, through various kind of instabilities (e.g., \citealt{Schure12} for a review). Indeed, particle-in-cell simulations (e.g., \citealt{Caprioli14}) can successfully demonstrate that starting with a flow with small magnetization, a quite strong field parallel to the shock develops close to the front as a result of the excitation of Alfven waves by streaming accelerated protons. The self-produced field dominates downstream in the close vicinity of the shock but decays quite rapidly due to various kinds of damping.

\citet{tavecchio20} presented a model for the X-ray polarization in HSPs based on a scenario assuming that the jet plasma is characterized by a small magnetization and a nearly parallel magnetic field (i.e., the field lines are almost parallel to the shock normal). As supported by the simulations mentioned above, a strong orthogonal field (i.e., orthogonal to the shock normal) component is assumed to develop close to the shock, decreasing with distance following a power law. The model further assumes that particles accelerated at the shock through DSA are advected downstream while cooling through the emission of synchrotron (and inverse Compton) radiation. High-energy electrons radiating in the X-ray band cool very rapidly, in a region dominated by the strong orthogonal self-produced field. Therefore, the model naturally predicts a large degree of polarization (around $40-50\%$) in the medium-hard X-ray band covered by IXPE. On the other hand, low-energy electrons, responsible for the optical-IR emission, travel over a much longer distance before completing cooling and therefore the resulting polarization, determined by the contribution of regions with field lines of different orientations, is strongly reduced (as observed, e.g., in \citealt{Pavlidou14}). 

A representative X-ray light curve predicted by this scenario is displayed in
Fig. \ref{mods.fig} (left panel). Here it is assumed that the shock is active for a finite time and therefore the injection of particles lasts for a given time $t_{\rm inj}$ (of the order of the light crossing time of the jet, assumed to have a radius of $r=10^{15}$ cm). With the assumed parameters (see \citealt{tavecchio20} for details) the almost constant polarization fraction predicted in the IXPE band is on the order of 40\%. 
An important point to note is that in this model, the essential constancy of the polarization angle at the highest energies is assumed (determined by the constant orientation of the self-produced fields). This, as shown below, has important consequences from the observational point of view.

\subsection{Acceleration via magnetic reconnection in kink-unstable jet}

Magnetized jets with an important toroidal component are subject to current-driven instabilities that, leading to the deformation of the jet, create conditions suitable for the reconnection of field lines and the dissipation of magnetic energy, a fraction of which can be tapped to accelerate non-thermal populations of particles.  

\citet{bodo21} derived the time-dependent emission quantities (including the polarization parameters) for a magnetized jet developing current sheets under the effect of a kink instability. Particles are injected in the simulation according to the current sheet properties and are followed while emitting and cooling. The polarization properties of the X-ray emission, produced by the rapidly cooling high-energy electrons, are mainly determined by the geometry of the magnetic fields close to the injection sites. Indeed, simulations show that the electrons are trapped inside the current sheets. 

The complex geometry of the current sheets and the associated magnetic fields inside the jet causes an effective dilution of the polarization, which, in fact, does not exceed 20\%. Moreover, the evolution of the structures drives a temporal modulation of the polarization properties. In Fig. \ref{mods.fig}, middle panel, we report a light-curve obtained from the simulation of \citet{bodo21} that we use for our study.The polarization angle light-curve was converted in the working range of our simulations, namely, -90\degr$-$+90\degr.

\subsection{Shock in a turbulent jet}

\begin{figure*}
\begin{minipage}[c]{0.95\textwidth}
 \includegraphics[width=0.32\textwidth]{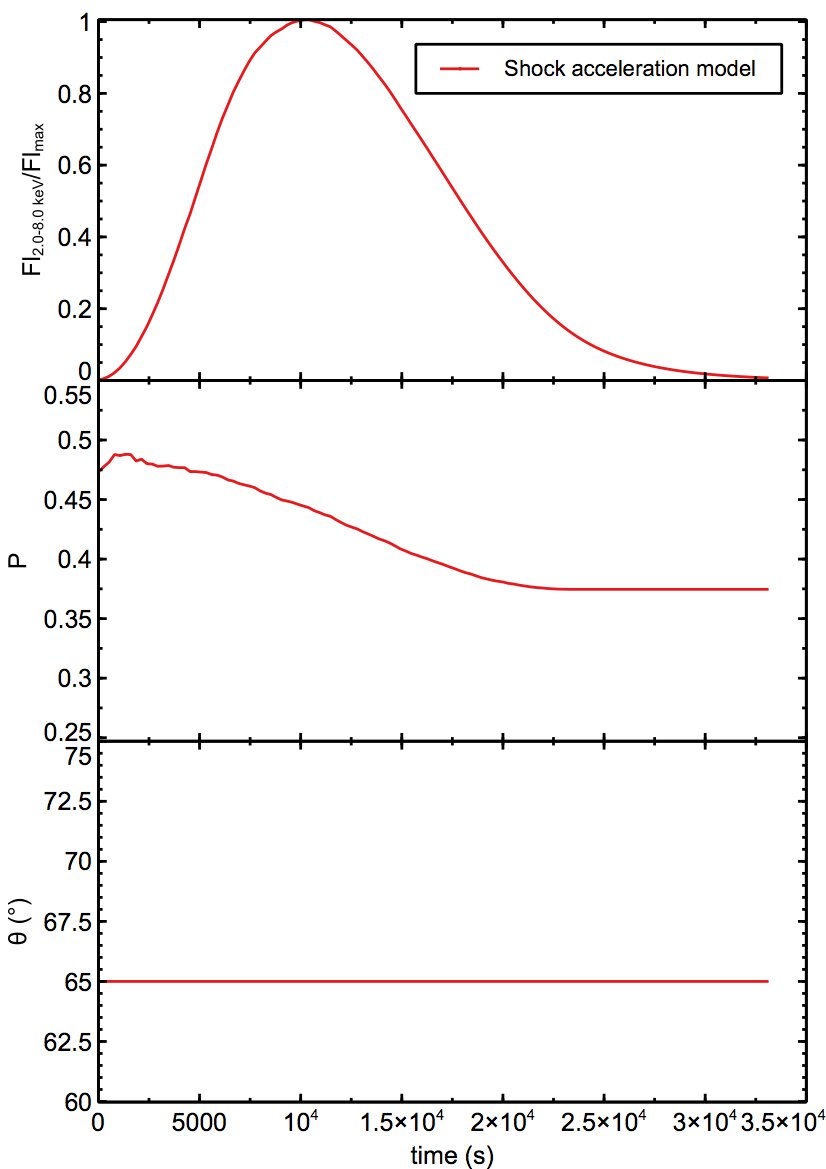}
 \includegraphics[width=0.32\textwidth]{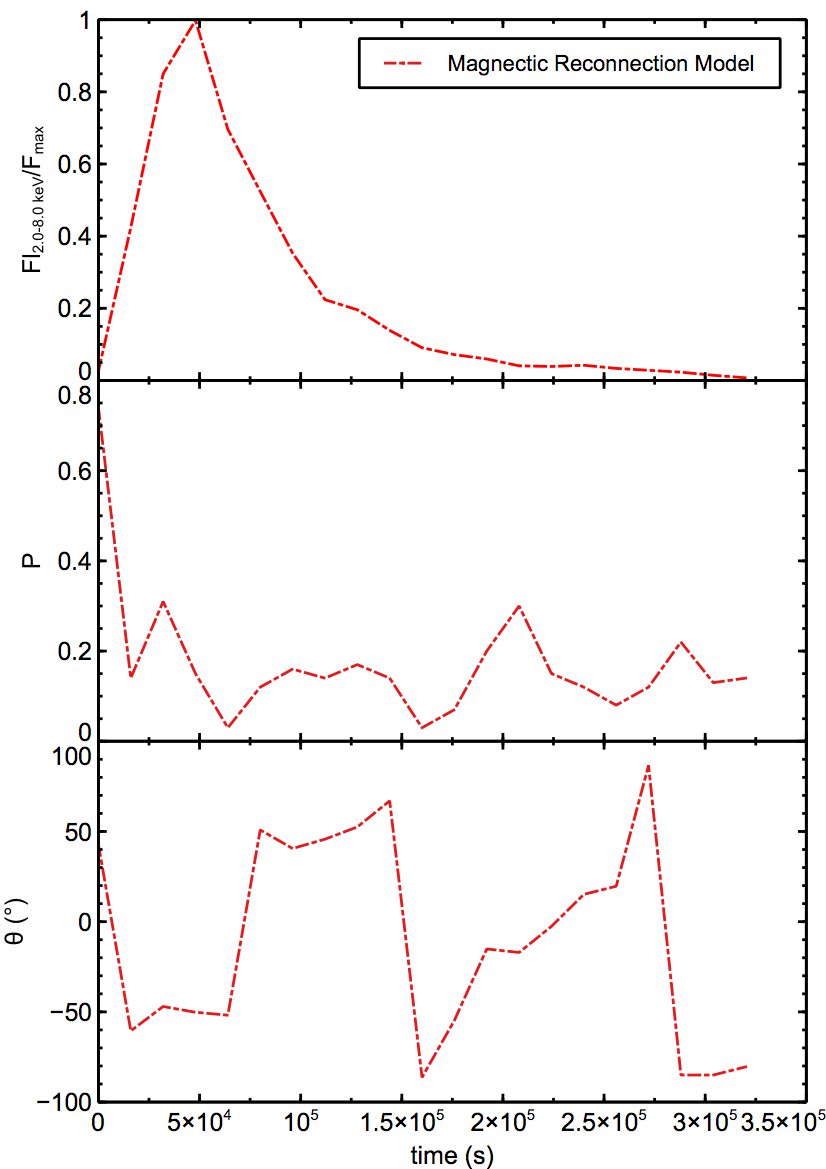}
 \includegraphics[width=0.32\textwidth]{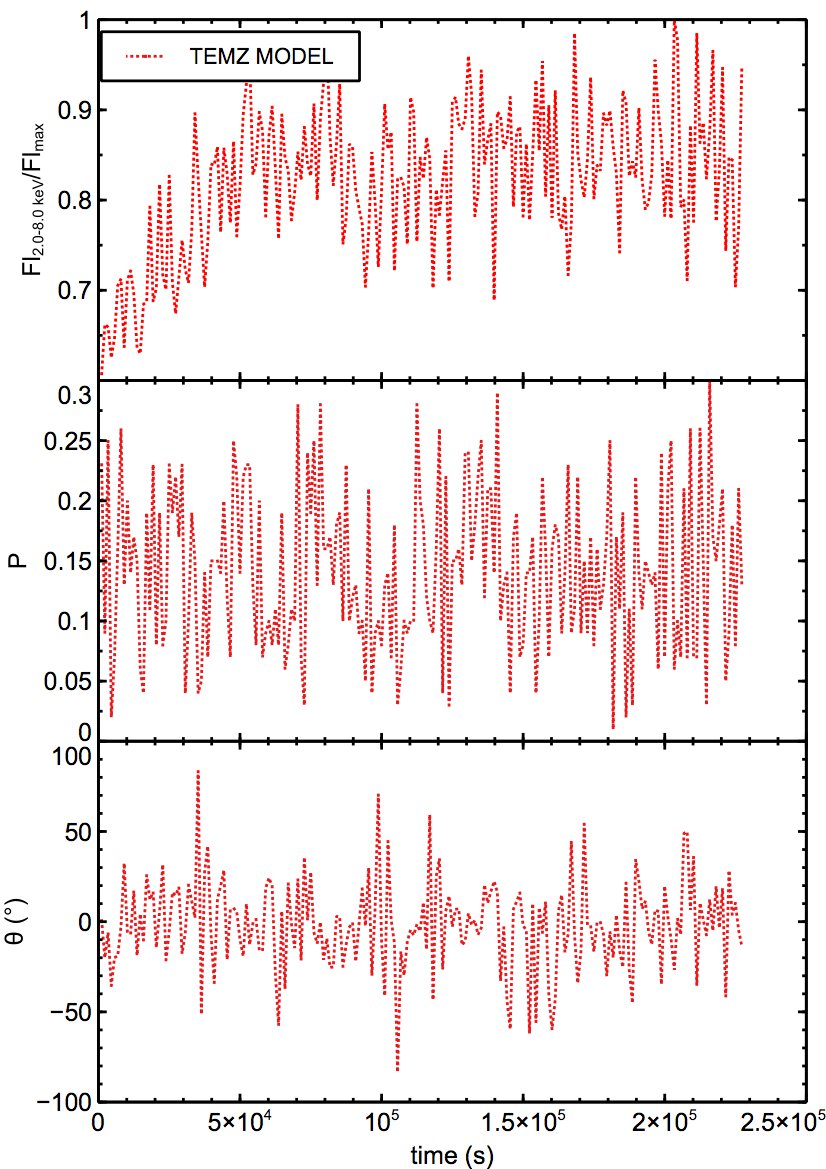}
\end{minipage}
 \caption{Input models for our simulations: shock acceleration model (left figure, solid line), magnetic reconnection model (middle figure, dashed-dotte line), and TEMZ model (right figure, dotted line). In each figure, from the top to the bottom, we show the time evolution of the flux, the polarization degree, and the polarization angle. The flux is normalized to the maximum value, for display purposes only.}
  \label{mods.fig}
\end{figure*}

Turbulence is a common phenomenon in hydrodynamical and MHD flows. Therefore, jets are expected to be turbulent, at least to some extent. The tangling of the magnetic field lines produced by turbulence might then have a major impact on the polarization properties of the synchrotron radiation. 
We consider here the specific turbulent model introduced by \citet{marscher14} as revised by \citet{marscher17} and \citet{marscher21}. In this framework, the emission region is proposed to be downstream flow beyond where the turbulent plasma crosses an oblique shock (resulting, for instance, from recollimation). Electrons accelerated at the shock emit synchrotron and inverse Compton radiation in the post-shock plasma where the magnetic field is partially ordered perpendicular to the shock normal by hydrodynamic compression at the shock front. For simplicity, turbulence is modelled by dividing the jet into nested zones within which the physical properties (electron distribution, magnetic field) are randomly selected but uniform. Relativistic electrons in the pre-shock flow are heated by compression in all of the cells. Furthermore, electrons in cells where the magnetic field is nearly parallel to the shock normal are accelerated to even higher energies via DSA. Therefore, fewer cells contain electrons with the higher energies needed to radiate at higher frequencies. A large number of cells can radiate at lower frequencies, so that the polarization is low from averaging of the random turbulent component. On the other hand, at higher frequencies, fewer cells are involved in the emission, so the mean polarization is higher, and it, as well as the polarization angle are more highly variable. In Fig. \ref{mods.fig} (right panel), a realization of the turbulent model in the IXPE band is shown.
We used these light curves in the simulations, as described in the following section.

%
%
%
\begin{figure}
\includegraphics[width=0.5\textwidth]{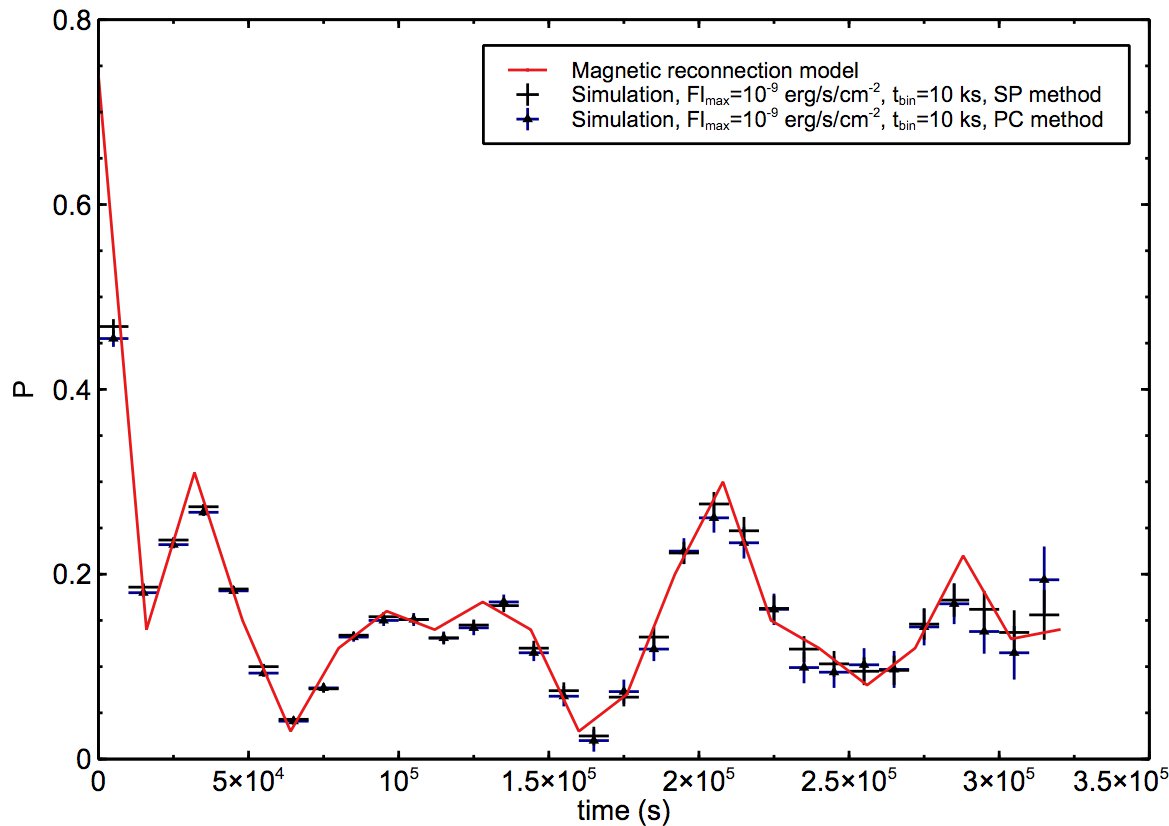}
\caption{Theoretical (solid line) and simulated (data points) polarization-degree light curves for the case of the magnetic reconnection model. The simulated data are from a run with $Fl_{\rm max}=10^{-9}$ \ergsc and $t_{bin}$=10 ks. We show that the PCUBE method (PC, blue triangles) and the spectropolarimetric fits (SP, green dots) returns consistent results.}
\label{cfr_met.fig}
\end{figure}


\section{Simulations}
We simulated IXPE observations of a variable blazar 
behaving as expected under the shock acceleration, magnetic reconnection, and TEMZ models presented above. Our study is aimed at determining the detectability of the polarization and of the temporal variability of the polarization properties characteristic for each model. We also studied how the detected polarization declines when  observations longer than the timescale of variability are performed. This informs future exploitation and interpretation of IXPE data.\\
We performed the simulations using the {\tt ixpeobssim} tool, version 18 \citep{pesce2019}. This is a Python-based framework that convolves the user defined source model, including spectral and polarization information with the IXPE instrument response functions (e.g., the point spread function of the telescopes and the effective area of the detectors). The spectral and polarization properties can be time-dependent, as prescribed by our cases of time variable blazars. The output of the simulations are simulated event files, which are treated with the same analysis tools of real observed data to create spectra and polarization products. In order to cope with the computational complexity of the problem, we used HTC-based high-end computers available on the Amazon Web Services Cloud Computing platform \citep{landoni}. In particular, we adopted instances with at least 32 vCPU (128 vCPU at the peak) with 128 GB of memory (1 TB at the peak). The adoption of Cloud Computing allowed us to quickly perform our simulation while reducing the cost of dedicated hardware.\\
In the polarization analysis, we used the Stokes parameters \citep{Stokes}
formalism, as implemented in \citet{Kislat2015}, to derive the polarization degree ($P$) and angle ($\theta$) from the Stokes parameters \I, \Q, and \U,\ where $\it I$ is the total intensity,  $\it Q$ is  the linearly polarized radiation intensity along the reference frame axes, and $\it U$ is the linearly polarized radiation intensity at $\pm$45\degr with respect to the main reference frame axis. The uncertainty in the determination of the polarization is quantified by the  minimum detectable polarization (\mdp\, \citealt{weisskopf2010}), which represents the minimum degree of polarization that can be determined with a 99\% probability against the null hypothesis, defined as:
\begin{equation}
        MDP = \frac{4.29}{\mu R_{\rm S}} \sqrt{\frac{R_{\rm S} + R_{\rm B}}{T}}
        \label{eq:mdp}
,\end{equation}
where $\it R_{\rm S}$ is the detected source rate (in counts/s), $\it R_{\rm B}$ is the background rate (in counts/s), $\it T$ is the observation time (in seconds) and $\it \mu$ is the adimensional modulation factor of the detector, namely, the response of the detector to 100\% polarized radiation at a given energy.  In practice, in case of a non-detection, the \mdp\ prescribes an upper limit for the observed polarization degree. In the case of a detection, the \mdp\ determines the error of the polarization degree \citep{Kislat2015} and, therefore, the statistical significance of the measurement.\\
We proceeded with the following steps. We simulated a point-like source, with a power-law spectrum with a photon index $\Gamma=2.0$. We set the normalization of the spectrum as time variable according to the flux light-curve of each of the three models (see Fig. \ref{mods.fig}). Next, we assigned the time-variable polarization degree and angle to the source. For the shock acceleration model, the polarization angle is constant over time. In the case of bright point-like sources such as blazars, the sky and instrumental background are negligible \citep[see e.g.,][]{DiGesu2020,ferrazzoli21}, thus, these parameters were not included
in our simulations.\\
We simulated observations lasting the entire time length of each light curve, namely, $\sim$30 ks, $\sim$350 ks, and $\sim$250 ks for the shock-acceleration, magnetic reconnection, and TEMZ model, respectively. The total length of the light curves is set by the theoretical models. Here, we are mainly interested in probing the viability of time-slicing the data to search for variability on a short time scale. For this, we used the {\tt xpselect} tool and we sliced the simulated event file into shorter exposed event files corresponding to time bins of equal length. In this way, we can test whether IXPE can retrieve the time variability of the polarization properties induced by the simulated models.\\
In each time bin, we determined the observed polarization degree, $P_{\rm obs}$ and angle  $\theta_{\rm obs}$, using two methods: a polarization cube and a spectropolarimetric fit. Here, the quoted polarization properties are always the mean over the determinations of the three IXPE DUs. A polarization cube is the simplest data structure holding polarization information and can be created using the PCUBE algorithm in {\tt ixpeobssim}. It applies the \citet{Kislat2015} formalism to a user-defined set of events to compute the Stokes parameters, the MDP, the polarization degree, the polarization angle, and their uncertainties. In our case, we created polarization cubes across the whole 2.0-8.0 keV energy band because, in our models, there are no substantial variations of the polarization properties with energy in this range. We carried out the instructive exercise of comparing the polarization degree and angle determined by the PCUBE algorithm with those obtained via a spectropolarimetric fit. For the spectropolarimetric fits, we use the {\tt ixpeobssim} interface to pyXSpec \citep{xspec}  and to the binning tool {\tt grppha}. We binned the event files into separated spectra for the $\it I$, $\it Q,$ and $\it U$ Stokes parameters, using the PHA1, PHAQ, and PHAU algorithms, respectively. Next, in order to perform a spectropolarimetric fit that meaningfully applies the $\chi^2$ statistic, we binned the \I\ spectrum requiring that a minimum of 30 counts is reached in each energy bin. We copied the same binning for the \Q\ and \U\ spectra. We fit the spectra with a model combining a powerlaw spectrum and a polarization degree and angle constant over the 2.0$-$8.0 keV energy band (powerlaw*constpol in XSpec syntax). From the fit in each individual time bin, we recorded the polarization degree and angle and their uncertainties.\\
In these ways, we produced observed light curves of the polarization degree
($P_{\rm obs}(t)$) and angle ($\theta_{\rm obs}(t)$). We find that the light curves obtained using the PCUBE  algorithm and the spectropolarimetric fits are always consistent with each other (see Fig. \ref{cfr_met.fig} as an example). This is a useful test of the readiness of the polarization analysis tools in light of the upcoming exploitation of IXPE data. In the following, unless otherwise stated, we show the light curves and results obtained via the PCUBE algorithm.\\
Using this procedure,we ran simulations varying the value of the maximum flux $Fl_{\rm max}$  for the each of the three models. We tested, when possible, three values of $Fl_{\rm max}$  1,5, and 10 $\times 10^{-10}$ \ergsc. These are plausible flux values for a typical IXPE target \citep{liodakis2019}. For each case of flux, we tested a variety of possible lengths of the time slices, in the range of 0.5$-$300 ks, depending on the model. In these tests we use a step of 5 ks for the shock acceleration and magnetic reconnection models, while for the TEMZ model we use a step of 1 ks. The aim is to determine, for each model and for each flux condition, which observing time bin best samples the variability pattern. Moreover, we can study how the polarization degree averages out when the time bins undersample the variability pattern. This is useful for the interpretation of real IXPE data.


\begin{figure*}
\begin{minipage}[c]{1.0\textwidth}
 \includegraphics[width=0.5\textwidth]{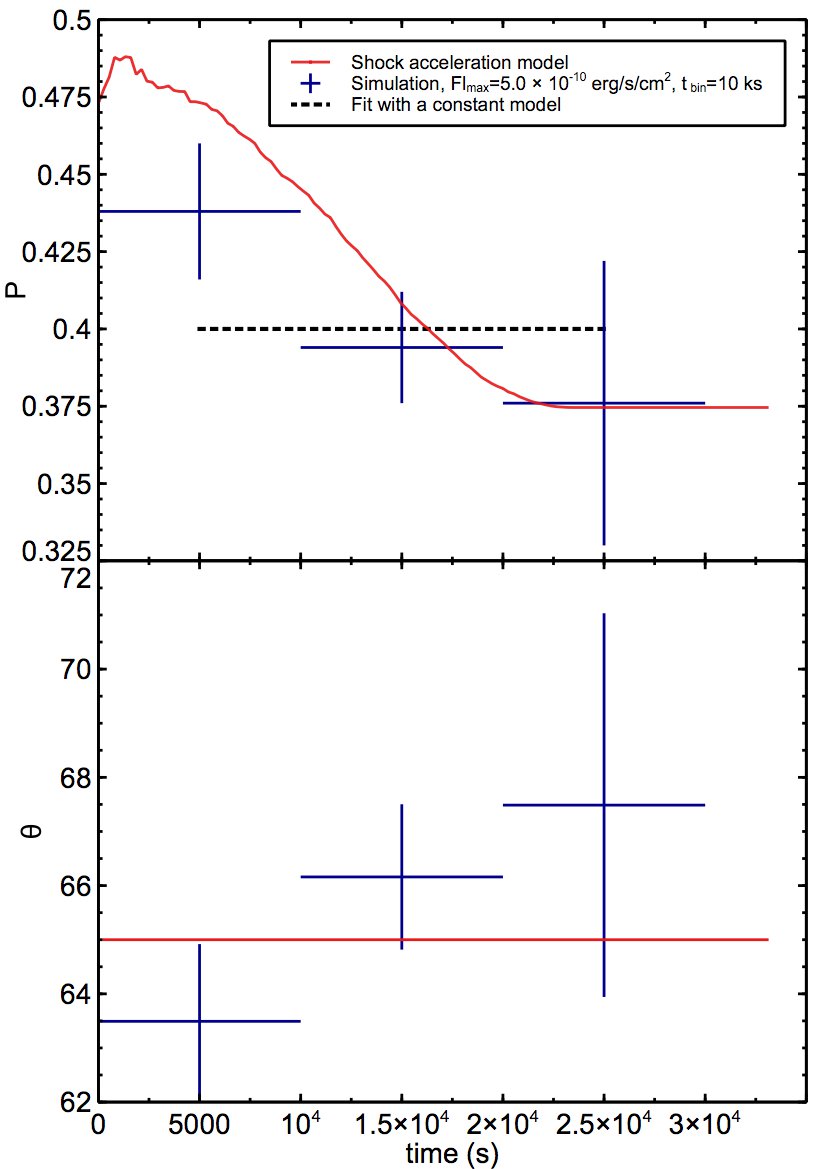}
 \includegraphics[width=0.5\textwidth]{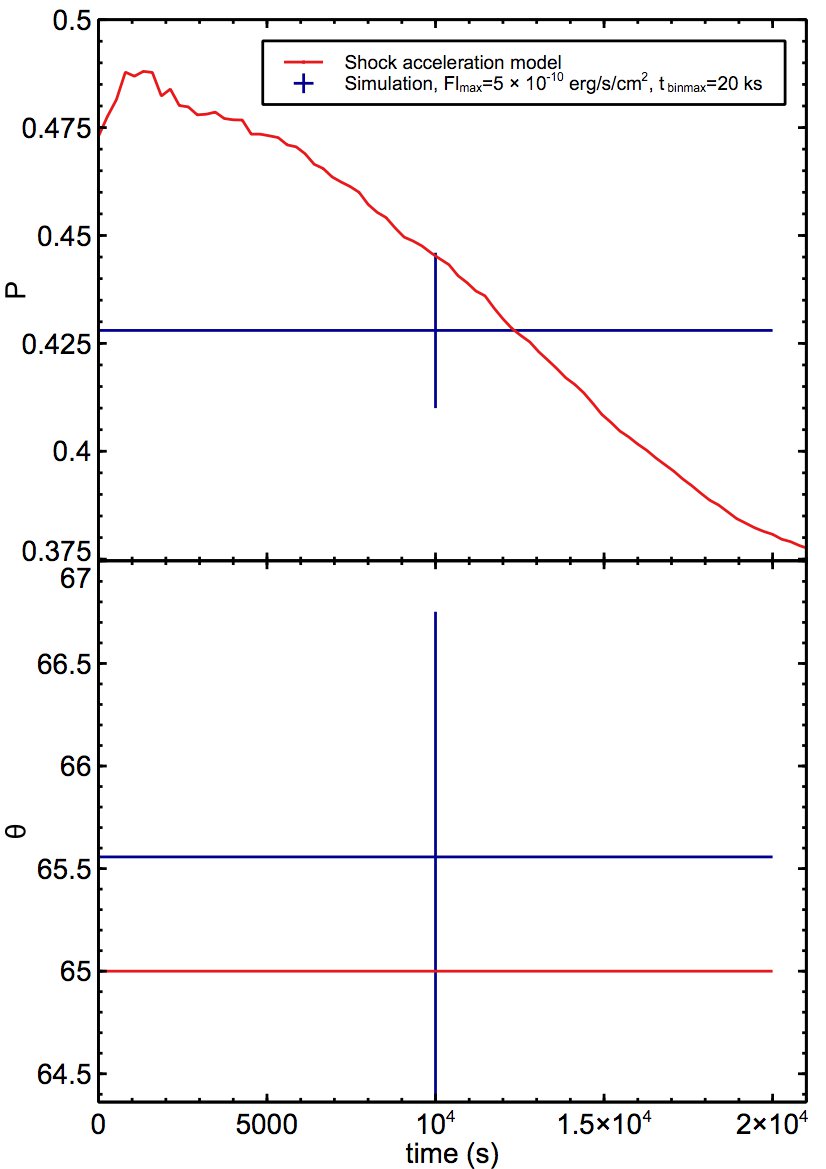}
\end{minipage}
 \caption{Results of the simulations with $Fl_{\rm max}=5 \times 10^{-10}$ \ergsc and $t_{\rm bin}=10$ ks (left figure), and  $t_{\rm binmax}=20$ ks (right figure) in the case of the shock acceleration model. In each figure, from the top to the bottom, we show the time evolution of the the polarization degree and polarization angle. In all the panels, the solid line represents the input model,  the data points represents the simulation results,
 while the dashed line represents a fit with
 a constant model.}
  \label{sims_tavecchio.fig}
\end{figure*}
\begin{figure*}
\begin{minipage}[c]{1.0\textwidth}
 \includegraphics[width=0.5\textwidth]{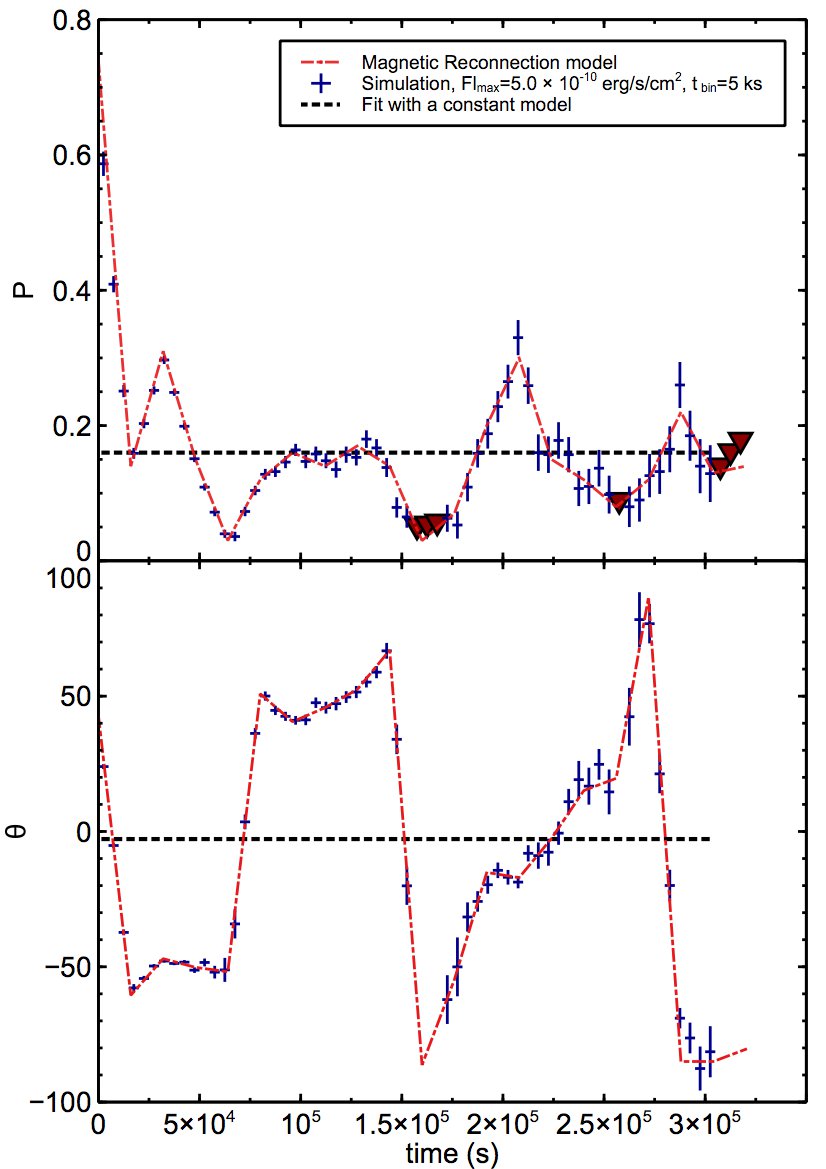}
 \includegraphics[width=0.5\textwidth]{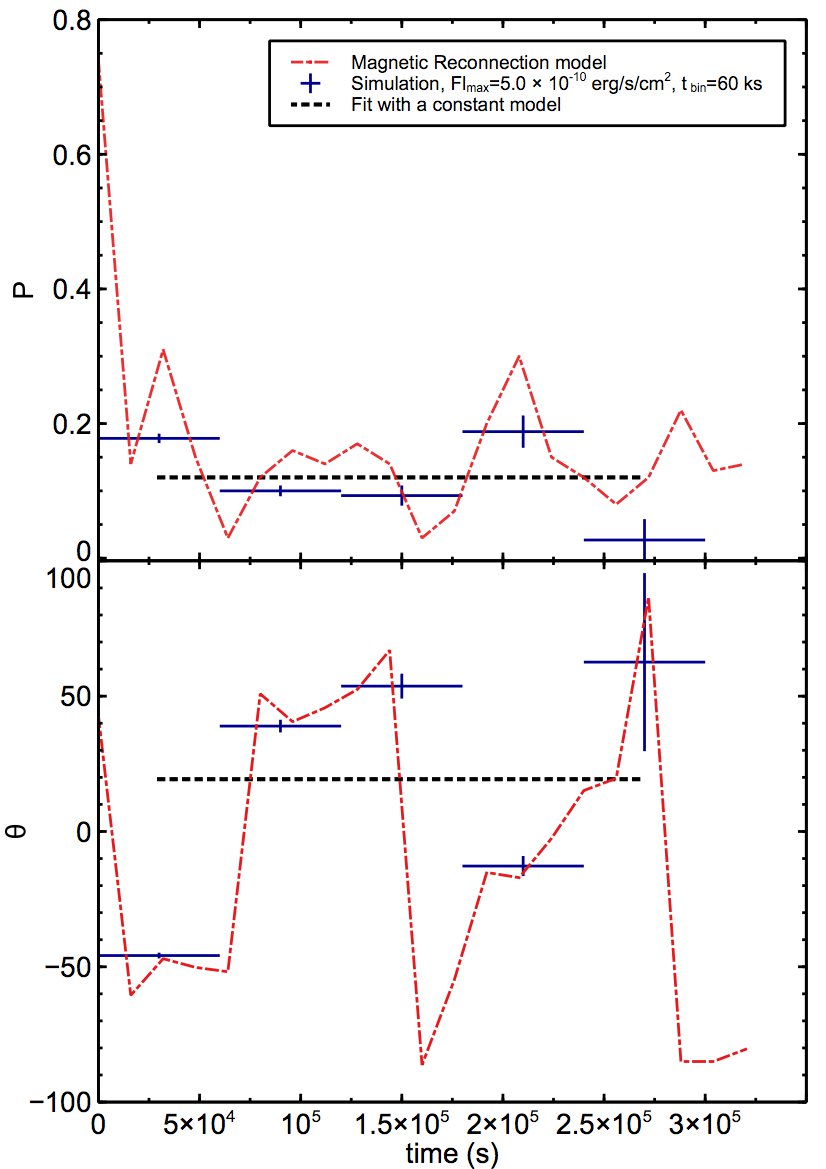}
\end{minipage}
 \caption{Results of the simulations with $Fl_{\rm max}=5 \times 10^{-10}$ \ergsc and $t_{\rm bin}=5$ ks (left figure) and  $t_{\rm binmax}= 60$ ks (right figure) in the case of the magnetic reconnection model. All the panels are the same as in Fig. \ref{sims_tavecchio.fig}. The dashed-dotted line represents the input model. The triangles are upper limits for the polarization degree. }
  \label{sims_bodo.fig}
\end{figure*}
\begin{figure*}
\begin{minipage}[c]{1.0\textwidth}
 \includegraphics[width=0.5\textwidth]{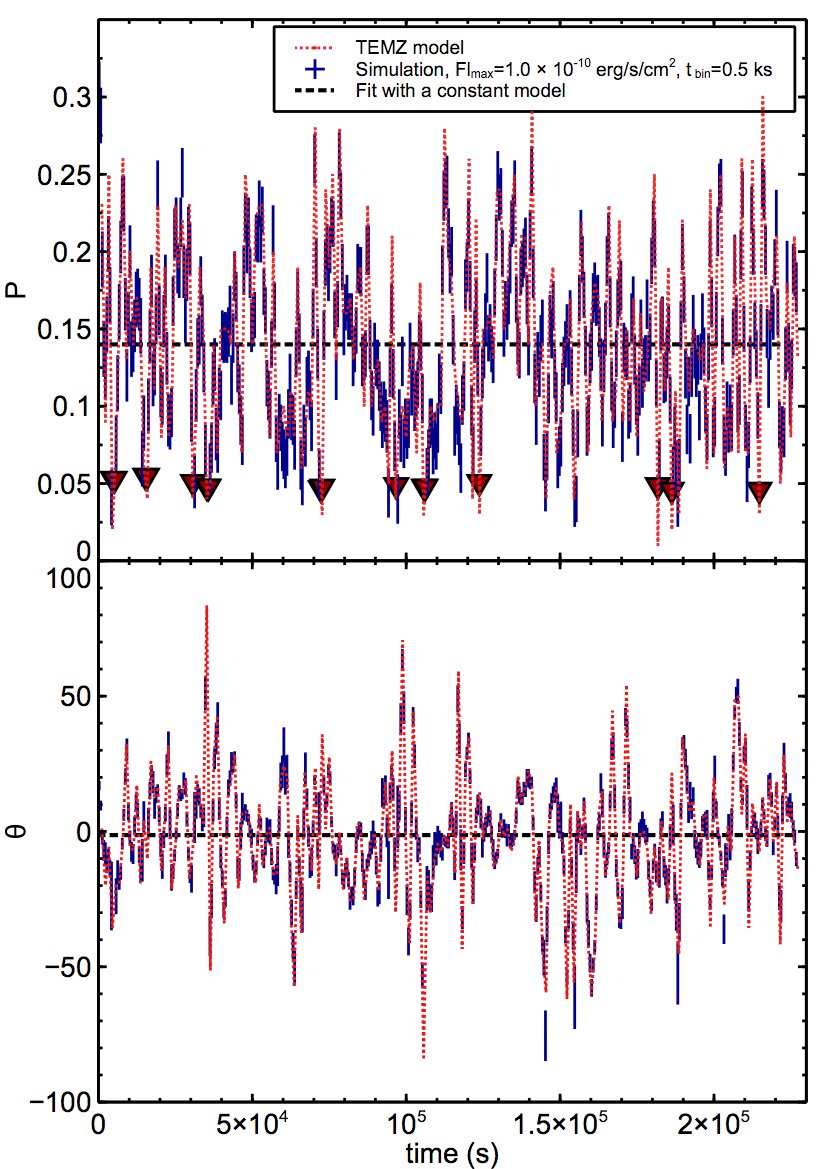}
 \includegraphics[width=0.5\textwidth]{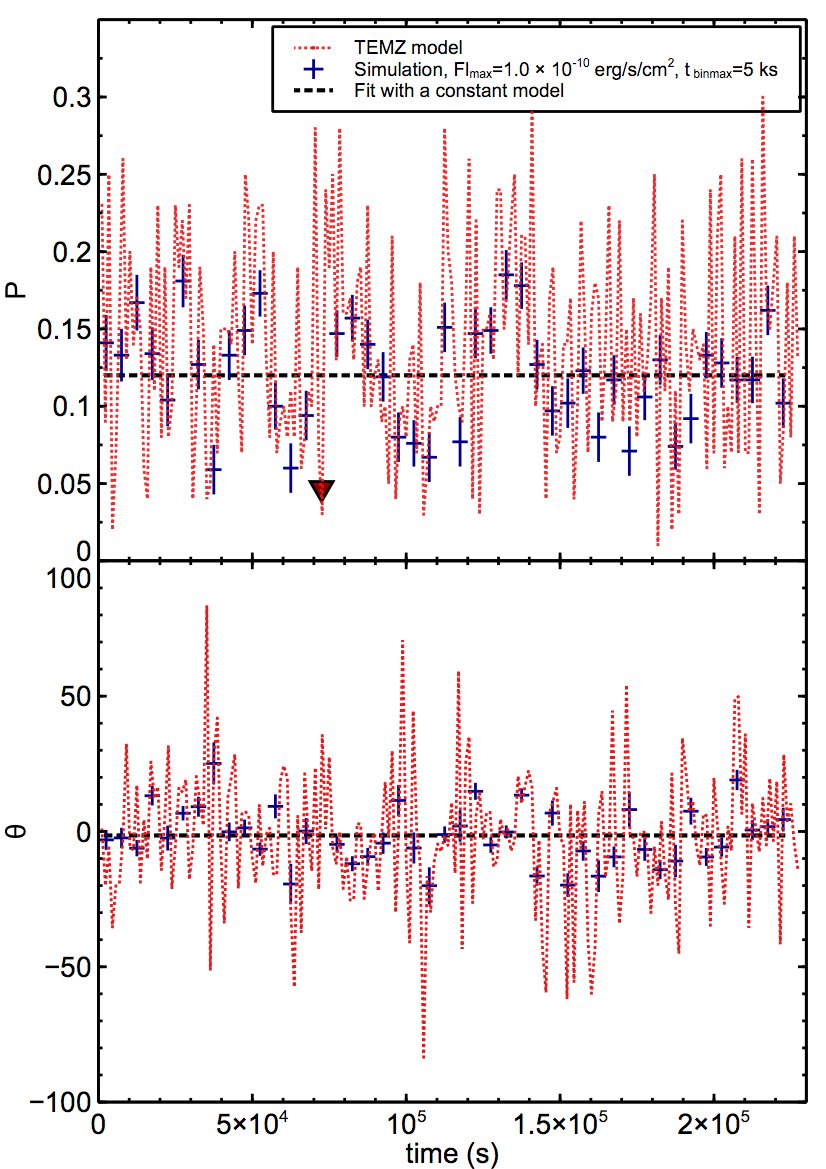}
\end{minipage}
 \caption{Results of the simulations with $Fl_{\rm max}=1 \times 10^{-10}$ \ergsc and $t_{\rm bin}=0.5$ ks (left figure) and  $t_{\rm binmax}=5$ ks (right figure) in the case of the TEMZ model. All the panels are the same as in Fig. \ref{sims_tavecchio.fig}. The dotted line represents the input model. The triangles are upper limits for the polarization degree.}
  \label{sims_marscher.fig}
\end{figure*}
%

\begin{table*}
\caption{Simulation results.}     
\label{sims.tab}      
\centering                    
\begin{tabular}{ccccc}        
\hline      
$Fl_{\rm max}$ \tablefootmark{a}&  
$t _{\rm bin}$ \tablefootmark{b} & $\chi^2_{\nu}(P)$ \tablefootmark{c} 
&$\chi^2_{\nu} (\theta)$ \tablefootmark{c}  \\
& $t _{\rm binmax}$ \tablefootmark{b}& &\\
( $10^{-10}$ \ergsc)& (ks) 
& MOD \tablefootmark{*} \quad CFIT \tablefootmark{**} 
& MOD \tablefootmark{*} \quad CFIT \tablefootmark{**} &\\
\hline
\multicolumn{3}{l}{Shock acceleration model \tablefootmark{***}} \\
\hline
\multirow{2}{*}{1.0} &10.0 & 4.6/3 \quad 0.12/2 & - \quad -\\
& -  & - \quad - & - \quad -\\ 
\hline
\multirow{2}{*}{5.0} & 10.0 & 3.22/3 \quad 3.20/2 & - \quad -\\
& -  & - \quad - & - \quad -\\  
\hline
\multirow{2}{*}{10.0} &10.0 & 5.06/3 \quad 7.59/2 & - \quad -\\
& -  & - \quad - & - \quad -\\ 
\hline
\hline
\multicolumn{3}{l}{Magnetic reconnection model}\\
\hline
\multirow{2}{*}{1.0} &10.0 & 27/17 \quad 246/16 & 25/18 \quad 4013/17 \\
& 70 & 113/2 \quad 0.46/1 &  \quad  0.46/2 \quad 345/1\\ 
\hline
\multirow{2}{*}{5.0} &5.0 & 82/57 \quad 3608/56 & 55/57 \quad 40318/56 \\
& 60 & 389/5 \quad 130/4 & 106/5 \quad 4041/4  \\ 
\hline
\multirow{2}{*}{10.0} &5.0 & 101/63\quad 7527/62 & 86/63 \quad 74172/62 \\
& 60 & 839/15 \quad 236/4 & 173/5 \quad 7941/4 \\ 
\hline
\hline
\multicolumn{3}{l}{TEMZ model \tablefootmark{****}}\\
\hline
\multirow{2}{*}{1.0} &0.5 & 471/438 \quad 4457/437 & 477/438 \quad 12862/437
 \\
& 5.0 & 455/44 \quad 202/43 & 602/44 \quad 281/43 \\ 
\hline
\multirow{2}{*}{5.0} & - & - \quad - & - \quad -  \\
& 5.0 & 2526/45 \quad 1170/44 & 2721/45 \quad 1481/44 \\ 
\hline
\multirow{2}{*}{10.0} & - & - \quad - & - \quad -  \\ 
& 5.0 & 5022/45 \quad 2419/44  & 5218/45 \quad 2833/44 \\ 
\hline
\end{tabular}
\tablefoot{
\tablefoottext{a}{Maximum value of the simulated flux, in 2.0-8.0 keV band.}
\tablefoottext{b}{Length of the time bin that we used in the simulation. For each model and for each case of flux, we report (when possible) the results for two cases of interest i.e. $t_{\rm bin}$ where the simulated data are best-fitted by the nominal input model and $t_{\rm binmax}$ where the best-fit constant model provides a \chirid\ lower than the that of the nominal input model.}
\tablefoottext{c}{Reduced $\chi^2$ obtained for models of the simulated $P_{\rm obs}(t)$.}
\tablefoottext{d}{Reduced $\chi^2$ obtained for models of the simulated $\theta_{\rm obs}(t)$.}
\tablefoottext{*}{Injected model.}
\tablefoottext{**}{Best-fit constant model.}
\tablefoottext{***}{For the shock acceleration model,
the case of $t_{\rm binmax}$ is not listed, because
it is a single time bin of 20 ks. We also did not list any $\chi^2 (\theta)$ because the polarization angle is not time variable in this model.}
\tablefoottext{****}{For the TEMZ model,
it was computationally prohibitive to run simulations with time bins lower
than 1 ks for the two higher values
of flux.}}
\end{table*}


\section{Results}
We show two examples of output of our simulation runs in Figs. \ref{sims_tavecchio.fig},\ref{sims_bodo.fig}, and \ref{sims_marscher.fig} for the shock-acceleration, magnetic reconnection, and TEMZ models, respectively. 
Unless otherwise stated, we took the case of $Fl_{\rm max}=5 \times 10^{-10}$ \ergsc\  as an example for display. In all figures, the observed light curves
$P_{\rm obs}(t)$ and $\theta_{\rm obs}(t)$ are visually compared with those of the input models $P_{\rm mod}(t)$ and $\theta_{\rm mod}(t)$. In those time bins where the polarization degree is undetected, we show the $3\sigma$ upper limit for $P_{\rm obs}$ prescribed by the nominal \mdp. Accordingly, the polarization angle cannot be determined and thus it is not shown.\\ 
To quantitatively assess the performance of the observations in detecting the time variability of the polarization properties, we used a $\chi^2$ test. In each simulation run, we computed the reduced $\chi^2_{\rm \nu,mod}$ relative to the input model and the reduced $\chi^2_{\rm \nu, cfit}$ relative to the best-fit constant model. We perform the fit of the observed light curves $P_{\rm obs}(t)$ and $\theta_{\rm obs}(t)$ with a constant model using the $curve\_fit$ function in Python. The reduced $\chi^2$ values are given, taking, for instance, the case of $P_{\rm obs}(t)$ and of the input model via:
$$
\chi^2_{\rm \nu, mod}=
\frac{\sum_{\rm tbins} \frac{P_{\rm obs}(t)-P_{\rm mod}(t)}{\sigma^2_{\rm P}}}{\rm d.o.f.}
$$
where the sum is intended over all the time bins, $P_{\rm mod}(t)$ is the value of the input model polarization degree in each time bin, $\sigma_{\rm P}$ is the error of the observed polarization degree in each time bin, and the degrees of freedom $\rm (d.o.f.)$ are given by the number of the time bins with detected polarization (in the case of the input model) minus one (in the case of the best-fit constant model).\\
The footprint of the shock acceleration mechanism is a high ($\sim$40\%) and quasi-constant polarization degree and a time constant polarization angle. Thus, in this case, provided that the polarization properties can be significantly measured, it is not critical to detect the time variability. We find that the polarization degree and angle would be generally be measured, in all the tested flux conditions and even with a single time bin, as 20 ks long. For instance, in the case shown in Fig. \ref{sims_tavecchio.fig}, right panel, we retrieve $P=42\pm2 \%$ and $\theta=66\deg \pm 1\deg$, which is consistent with the input model, where $P$ varies between $\sim 37\%$ and $\sim 48\%$. We still tested the possibility of detecting the time variability of the polarization degree when using smaller time bins. As we report in Table \ref{sims.tab}, for all the tested flux conditions, we find that the injected model provides the best fit to the data
when using time bins of 10 ks (Fig. \ref{sims_tavecchio.fig}, left panel). We also find that with this time slicing, a constant model is rejected on the basis of a formal $\chi^2$ test.\\
For the case of the magnetic reconnection and TEMZ models, it is critical to detect the time variability of the polarization properties, as they are a specific observable consequence of the proposed acceleration mechanisms. For these two models, we report two cases of interest in Table \ref{sims.tab} and Figs. \ref{sims_bodo.fig} and \ref{sims_marscher.fig}. We label as $t_{\rm bin}$ the case where the injected input model provides the best fit (i.e., the minimum  $\chi^2$ and p-value observed in our tests) to the data, indicating that the observations are capable of following the time variability of the model. As we increase the length
of the time slices, we find that the $\chi^2$ of the fits of the simulated light curves with the nominal input models increases. We define as $t_{\rm binmax}$ the time bin after which the $\chi^2_{\rm \nu, cfit}$ provided by the best-fit constant model becomes lower than the $\chi^2_{\rm \nu, mod}$ provided by the nominal input model. Although the constant model never provides a statistically acceptable fit to the data, the case of $t_{\rm binmax}$ is a condition in which it is not possible to assert that the input model fits the data better than a constant, that is, the data are inconclusive in terms of discriminating the time variability of the model.
For the magnetic reconnection model, we find a $t_{\rm bin}$  of 5.0-10.0 ks, depending on the value of $Fl_{\rm max}$ (Table \ref{sims.tab}). In this condition, the observed data points are in good agreement with the theoretical light curves of the polarization degree and angle (Fig. \ref{sims_bodo.fig}, left panel). Under this condition, the $\chi^2$ values clearly disfavor the fit with a constant model (see Table \ref{sims.tab}). Conversely, $\chi^2_{\rm \nu, cfit}$ becomes inferior to $\chi^2_{\rm \nu, mod}$ for time bins of 60-70 ks, again depending on the value of $Fl_{\rm max}$. This happens only for the simulated polarization degree time series, while in the case of the polarization angle time series we still find $\chi^2_{\rm \nu, mod} < \chi^2_{\rm \nu, cfit}$, likely because of the more prominent variability in the case of the polarization angle. When the capability of following the time variability is lost, we would still be able to measure a polarization degree, albeit diluted, because of the mixing of the different values along the light curve. For instance, in the first two time bins of Fig. \ref{sims_bodo.fig} (right panel), the polarization degree is measured with a relative uncertainty of 4\% and 7\%, while the polarization angle is retrieved with an error of a few degrees. The measure becomes more uncertain in the last two time bins (i.e., up to a relative uncertainty of 70\% for the polarization degree), because of the worsening of the statistic as the flux decreases along the light-curve.\\
The case of the TEMZ model (Fig. \ref{sims_marscher.fig}) is more demanding for the diagnostic capability, because of the rapid variability of both polarization degree and angle. Indeed, for this model, we found that $t_{\rm bin}$=0.5 ks is needed to obtain a formally good fit of the observed light curves with the injected model (Table \ref{sims.tab}). We could perform the test with this small time bin only for the case of $Fl_{\rm max}$=1$\times 10$ \ergsc\
(Fig. \ref{sims_marscher.fig}, left panel), because it was computationally prohibitive to run this simulation case for higher values of $Fl_{\rm max}$. Thus, for the higher values of $Fl_{\rm max}$, we list in Table \ref{sims.tab} only the case of $t_{\rm binmax}$, where $\chi^2_{\rm \nu, cfit}$ becomes inferior to $\chi^2_{\rm \nu, mod}$. We find for all the tested flux conditions and for both the polarization degree and polarization angle time series, a $t_{\rm binmax}$ length that is much shorter than in the previous case, namely, 5 ks (e.g., Fig. \ref{sims.tab}, right panel).\\
As a final remark, we note that Figs. \ref{sims_tavecchio.fig}-\ref{sims_marscher.fig} illustrate how observing a blazar variable in polarization for an observing time that is too long compared to the time variability of the source results in the depolarization of the signal. However, here we find that for realistic particle acceleration mechanisms and flux conditions, this effect does not make the polarization degree undetectable by IXPE.

\section{Summary and discussion}
The time variability of the X-ray polarization properties of an HSP blazar informs us about which mechanism is accelerating the particles in the jet. Potential in situ physical processes, such as diffusive shock acceleration or magnetic reconnection in a kink unstable jet, are expected to produce a distinctive time variability pattern in the X-ray polarization. Modeling of the shock acceleration scenario predicts a large ($\sim$40\%), quasi-constant polarization degree and constant polarization angle. In the magnetic reconnection scenario, a lower polarization degree is expected (below $\sim$20\%), with both the polarization degree and angle modulated with time. Turbulence in the jet may also influence the expected X-ray polarization, lowering the polarization degree and producing rapid variability in both the polarization degree and angle.\\
Thanks to the IXPE satellite, it will be possible, for the first time, to measure the X-ray polarization of extragalactic sources such as HSP blazars. Thus, we will have the unprecedented possibility of directly comparing the prediction of different models for particle acceleration in the jet with the observations. Observations of sources rapidly variable in polarization are potentially challenging because the short observing times needed to sample the variability could result in an unsignificant measurement of the polarization degree, if the collected photon counts are insufficient. On the other hand, the polarization degree may average to lower values when integrated over a a time bin that undersamples the intrinsic variability of the source.\\
In this work, we have presented simulations of IXPE observations of a HSP blazar variable in X-ray polarization. Our simulation work is fed by the theoretical light curves of the X-ray polarization degree and angle of \citet{tavecchio20}, \citet{bodo21}, and \citet{marscher21} for the shock acceleration, magnetic reconnection, and TEMZ model,
respectively.
Thereby, we tested how IXPE will be able to address the issue of particle acceleration mechanism in HSP blazars using physically motivated models and realistic IXPE data products and analysis techniques.\\
Our exercise demonstrates that time-slicing a long-exposure IXPE event file into shorter-exposure event files corresponding to time bins of arbitrary lengths is a viable data analysis strategy when searching for a variability in the polarization properties of a source. This indicates how we can treat real data when there is a reasonable expectation of time variability in the polarization of the source. Moreover, our simulation exercise offers the possibility of testing the methods of recovering the polarization degree and angle, namely, the polarization cube and the spectropolarimetric fit. We find (Fig. \ref{cfr_met.fig}) that both methods are robust in retrieving the polarization properties in our case where the polarization degree and angle are constant with energy.\\
Besides providing a probe of the data analysis techniques, our simulations provide useful indications of the conditions in which IXPE data permit us to detect the variable polarization of a blazar and to discriminate between different time variability patterns. First and foremost, we found that for all the three models and flux conditions (i.e., 1, 5, and 10 $\times 10^{-10}$ \ergsc) tested here, IXPE will generally be able to significantly measure some degree of polarization, even if diluted in an observation longer than the characteristic timescale of polarization variability. This is illustrated, for instance, in the right panels of Fig. \ref{sims_tavecchio.fig}, \ref{sims_bodo.fig}, and \ref{sims_marscher.fig} and we checked that this is true even when the light curves are compressed into a individual time bins.\\
By computing the $\chi^2$ of the fits of the simulated light curves with the nominal injected model and a best-fit constant model, we tested how well IXPE data probe the time modulation induced by each model. In the case of the shock-acceleration model (Fig. \ref{sims_tavecchio.fig}), the time modulation of the polarization degree is best fitted by the injected model (e.g., \chirid=1.07 for 3 d.o.f, see Table \ref{sims.tab}) by IXPE data taken in time bins of $\sim$10 ks. However, the key observable of this model is the high ($\sim$ 40\%) and quasi-constant polarization degree and angle, rather than the mild time modulation. We find that in all the tested flux conditions, a polarization degree consistent with the average of the light-curve will always be measured, even when using an individual time bin of 20 ks.
In the case of the magnetic reconnection model (Fig. \ref{sims_bodo.fig}), we were able to best fit (e.g., \chirid($P$)=1.43 and \chirid($\theta$)=0.96 for 57 d.o.f in Table \ref{sims.tab}) the simulated light curves with the nominal magnetic reconnection model when a series of IXPE observations 5-10 ks long was used. On the other hand, we find that for time bins of 60-70 ks, we are dealing with a condition for which it is not possible to determine whether the magnetic reconnection model fits the data better than a constant.\\
Our magnetic reconnection time sample displays a smooth polarization angle swing occurring on a time scale of hundreds of ks. However, polarization angle swings are possible even on shorter time scales, for instance, as predicted by recent particle in-cell simulations of magnetic reconnection
\citep{zhang2018}.
To see whether IXPE can track fast polarization angle swings, we made a simulation run for our magnetic reconnection sample with the time axis compressed by a factor of ten (i.e., for a total flare duration of $\sim$30 ks). Thus, this mimics the case of a polarization angle swing occurring intraday. We find that when the peak flux of the flare is the brightest of our tests (1$\times 10^{-9}$ \ergsc), a polarization angle swing is detected using time bins of 5/10 ks (e.g., we find
$\theta_{\rm obs}=-46\degr \pm 4\degr$ at t=5 ks and
$\theta_{\rm obs}=54\degr \pm 6\degr$ at t=15 ks). Thus, when favourable flux conditions are met, intraday polarization angle swings are detectable by IXPE.\\
Finally, in the case of the TEMZ model (Fig. \ref{sims_marscher.fig}), a more rapid variability of the polarization degree and angle is anticipated and, thus, it is critical to have the capability of collecting enough counts in short time bins, ideally shorter than one ks. Indeed, for this case, we found (Table \ref{sims.tab}) that the (statistically) best fit (\chirid($P$)=1.08 and \chirid($\theta$)=1.09 for 477 d.o.f.) of the simulated light curves with the nominal injected model is obtained when using time slices of 0.5 ks. Because of computational constraints,
we tested this only for the case of $Fl_{\rm max}$=1$\times 10^{-10}$ \ergsc.
For all the tested cases of $Fl_{\rm max}$, the nominal $\chi^2$ of a fit with constant model becomes lower than that that of the injected model when using time bins much shorter than in the previous case, namely, of 5 ks.\\
Overall, our exercise indicates that provided that the statistics of the observation allows for the data to be sliced into adequately short time bins, it is, in principle, possible to use IXPE to retrieve the time modulation of the polarization properties induced by popular models for particle acceleration in the jet. We note that the flux conditions that we tested here are a reasonable expectation \citep{liodakis2019} for the nearby HSPs that will be targeted by IXPE. \\
However, an obvious difficulty in interpreting real data is that it is not possible to perform analytical fits of the time series with different particle acceleration models, as we did here with  \virg{ad hoc} simulated data, because the models are not provided as grid models with adjustable parameters. Thus, our simulation exercise is useful also because it allows
for an estimation of the observable characteristic (e.g., the variance of the time series) that can be expected in the data when different physical mechanisms are in place. For instance, when we simulate the time constant polarization angle in the case of the shock acceleration model, using time bins of 10 ks, we find a standard deviation $\sigma$ of the time series of 1$\deg$-5$\deg$, depending on the flux. This is one order of magnitude lower than what we find for the polarization angle time series simulated assuming magnetic reconnection (i.e., $\sigma$=40$\deg$-48$\deg$, when using time bins of 10 ks) and still
lower than what we find for the TEMZ model (i.e.,
$\sigma$=7$\deg$-8$\deg$). This indicates that an IXPE polarization angle time series with a scatter on the order of some tens of degrees is difficult to reconcile with the case of the constant polarization angle predicted in the case of shock acceleration. Discriminating between the time variability induced by magnetic reconnection and turbulence in the flow can be more challenging. Here, we have shown that a pure magnetic reconnection tends to induce a greater degree of scatter in the polarization light curves and smooth variability that remains observable even when using time bins of some tens of ks.\\
Observations in optical polarization simultaneous to the IXPE pointing are also useful for the task of determining which mechanism is accelerating the particles in the jet. Indeed, the shock acceleration scenario of \citet{tavecchio20} predicts that while the X-ray polarization remains high and constant, the optical polarization decreases rapidly with both time and wavelength, because the optically-emitting particles cool slowly and have the time to travel farther from the shock region dominated by the self-generated magnetic field. Quite a different behavior is predicted in the case of magnetic reconnection and TEMZ \citep{bodo21, marscher14}, where the optical polarization is expected to display the same (smooth or erratic) modulation of the X-ray polarization; even though, in the TEMZ model, the optical variations are somewhat smoother than those at X-ray energies because lower electron energies, and therefore more cells, are involved). \\
A caveat of our models
for shock acceleration and magnetic reconnection is that they describe a possibly compact, upstream emission region that does not experience much turbulence. Several hybrid scenarios are possible and can be probed using IXPE data in combination with optical polarization data. For instance, finding that the X-ray polarization in HBL is low (e.g., $\sim$ 10\%) and similar to the optical would conform more to a TEMZ model where the optical and X-ray emission region are extended and partially cospatial. On the opposite extreme, finding a fairly high X-ray polarization in combination with a low optical polarization would point to a compact X-ray emission region close the the shock front with an optical emission region extending more downstream and being more affected by turbulence. Results that are in between these limits can be explained by different levels of a steady, turbulent component diluting the emission from the shocked region \citep{tavecchio20}. In practice, optical polarimetry can indicate the level of variability of the polarization of the synchrotron radiation of a blazar. Rapid, high-amplitude variability would conform more with reconnection and turbulent models. In addition, the combination of X-ray and optical polarimetry can probe how the X-ray and optically emitting particles are distributed in the jet and whether they experience the same level of turbulence.\\
In conclusion, we have shown that IXPE observations of nearby HSPs have the potential to provide unprecedented insights into the physics of particle acceleration in blazar jets. We used realistic data products and data analysis techniques to provide estimations of the
observable quantities induced by current theoretical models. Therefore, our simulation exercise informs the physical interpretation of IXPE data that are soon to become  available.

%

\begin{acknowledgements}
The Italian contribution to the IXPE mission is supported 
by the Italian Space Agency through agreements ASI-INAF n.2017-12-H.0
and  ASI-INFN n.2017.13-H.0.
FT acknowledges contribution from the grant INAF Main Stream project \virg{High-energy extragalactic astrophysics: toward the Cherenkov Telescope Array}.
The research of APM was supported in part by US National Science Foundation grant AST-2108622.
We thank Luca Baldini for helping with the ixpeobssim simulations.
\end{acknowledgements}


\end{document}